\documentclass[aps,pre,reprint,superscriptaddress,footinbib,floatfix]{revtex4-2}

\usepackage{graphicx}
\usepackage{bm}
\usepackage{amssymb}
\usepackage{amsbsy}
\usepackage{amsmath}
\usepackage{mathtools}
\usepackage{xcolor}
\usepackage{stmaryrd}
\usepackage{mathrsfs}
\usepackage[mathscr]{euscript}
\usepackage{tikz}
\usetikzlibrary{shapes}
\usepackage{cancel}
\usepackage[normalem]{ulem}

\begin{document}

\title{Dynamic Clustering of Active Rings}

\author{Ligesh Theeyancheri}
\affiliation{Department of Chemistry, Indian Institute of Technology Bombay, Mumbai 400076, India}
\author{Subhasish Chaki}
\affiliation{Department of Chemistry, Indian Institute of Technology Bombay, Mumbai 400076, India}
\affiliation{Institut f\"ur Theoretische Physik II - Soft Matter, Heinrich-Heine-Universit\"at D\"usseldorf, Universit\"atsstra\ss{}e 1, D-40225 D\"usseldorf, Germany}
\author{Tapomoy Bhattacharjee}
\email{tapa@ncbs.res.in}
\affiliation{National Centre for Biological Sciences, Tata Institute of Fundamental Research, Bangalore 560065, India}
\author{Rajarshi Chakrabarti}
\email{rajarshi@chem.iitb.ac.in}
\affiliation{Department of Chemistry, Indian Institute of Technology Bombay, Mumbai 400076, India}

\date{\today}

\begin{abstract}
\noindent A collection of rings made of active Brownian particles (ABPs) for different packing fractions and activities is investigated using computer simulations. We show that active rings display an emergent dynamic clustering instead of the conventional motility-induced phase separation (MIPS) as in the case of collection of ABPs. Surprisingly, increasing packing fraction of rings exhibits a non-monotonicity in the dynamics due to the formation of a large number of small clusters. The conformational fluctuations of the polymers suppress the usual MIPS exhibited by ABPs. Our findings demonstrate how the motion of a collection of rings is influenced by the interplay of activity, topology, and connectivity.
\end{abstract}

\maketitle

\section*{I\lowercase{ntroduction}}

\noindent Active agents such as bacteria~\cite{wu2000particle}, colloids locally driven by chemical reactions~\cite{illien2017fuelled,samin2015self,buttinoni2012active}, temperature gradient~\cite{jiang2010active} or by some other means operate far from thermal equilibrium as they use an internal source of energy to generate directed motion~\cite{bechinger2016active} and break the time-reversal symmetry~\cite{o2022time}. Biological examples of collection of active agents include suspensions of motile microorganisms, collective organization of cells in living tissues, and flocks of birds~\cite{peruani2006nonequilibrium,ramaswamy2003active,gopinath2012dynamical,redner2013structure,
vicsek2012collective,grober2023unconventional,garcia2015physics,hopkins2023motility}. Synthetic examples include vibrated granular monolayers~\cite{deseigne2010collective} and suspension of phoretic colloidal particles~\cite{ginot2015nonequilibrium}, etc. Collectively these active units show interesting behavior such as swarming~\cite{ramaswamy2003active,grober2023unconventional}, flocking~\cite{ramaswamy2010mechanics}, and nonequilibrium ordering~\cite{prathyusha2018dynamically}.\\

\noindent Another interesting phenomena has been observed for a collection of active particles is motility-induced phase separation (MIPS) where the system phase separates into two distinct phases, dilute and dense~\cite{redner2013structure,cates2015motility}. This is intriguing as it is purely due to persistent active motion in the absence of any sort of attractive interaction~\cite{speck2014effective,basu2018active}. How the phase behavior is affected by the polydisperse nature of the active particles~\cite{redner2013structure,cates2015motility,vicsek2012collective}, and the softness of interactions~\cite{de2022motility,martin2021statistical} have also been studied. The majority of the studies on MIPS have focused on active colloids that lack conformational fluctuations of the system. Recent works on complex active systems, like active polymers, have shown rich and counter-intuitive structural and dynamical properties~\cite{eisenstecken2016conformational,mousavi2019active,ghosh2014dynamics, liverpool2001viscoelasticity,roichman2007anomalous,duman2018collective,
rosa2014ring,nahali2016density,choi2021relative,wen2023collective,goswami2022reconfiguration,
gnan2019microscopic,smrek2020active,chattopadhyay2023two,venkatareddy2023effect,huang2023bridging,miranda2023self,locatelli2021activity}. The coupling of activity and conformations of polymers give rise to novel phenomena, such as an activity-induced polymer collapse or swelling and a transition from a trapped state to escape in porous media~\cite{theeyancheri2022migration,chopra2022geometric,theeyancheri2023active}. This illustrates that the understanding of active processes in the collection of polymers may be essential to designing new classes of active soft materials.\\

\noindent Our goal in this work is to study the dynamics and structure of a collection of self-propelled disks (in 2D) connected by springs to form a ring made of active disks. In the absence of these springs, the system undergoes MIPS depending on density and activity~\cite{redner2013structure,cates2015motility}. But will these active rings also separate into a dense and dilute phase, in other words, undergo MIPS? This remains unexplored to the best of our knowledge. However, if the system does not show MIPS, what is the underlying reason, and if it does, then how does it differ from the usual MIPS as shown by ABPs? This is an important question to ask, as polymers undergoing phase separation due to some sort of motility has been a topic of research in recent years~\cite{smrek2020active,agrawal2017chromatin,shi2018interphase,patra2022collective}. This is pertinent in the context of chromatin~\cite{agrawal2017chromatin,shi2018interphase}, malaria parasites~\cite{patra2022collective}, etc. For example, the chromatin phase separates into dense and dilute phases~\cite{agrawal2017chromatin,shi2018interphase}. The activity, in that case however, was introduced by putting a fraction of consecutive monomers on a polymer to stronger thermal fluctuations than the rest, i.e., by making them effectively hotter.\\

\noindent Our simulations with active rings show that the dynamics is strikingly different for rings compared to that of ABPs. While ABPs undergo MIPS, rings made of ABPs do not, instead formation of intermittent unstable clusters is observed. These clusters are ``dynamic'' as they form and disintegrate continuously. Further analyses of cluster size and numbers reveal that increasing activity (active force on each monomer) facilitates the formation more number of smaller clusters. Notably, faster dynamics of this collection of active rings emerge at higher packing fractions due to the combined effects of the activity-induced shape deformation and dynamic clustering of the rings. The formation of highly motile and dense small clusters at higher $\phi$ is due to the interplay of activity, deformation, and stress governing events in the system. Our model demonstrates the emergence of a fascinating dynamic clustering of active rings made of ABPs, which is different from the conventional MIPS behavior exhibited by ABPs.

\section*{M\lowercase{odel and} S\lowercase{imulation} D\lowercase{etails}}

\begin{figure}[h!]
\centering
\includegraphics[width=0.99\linewidth]{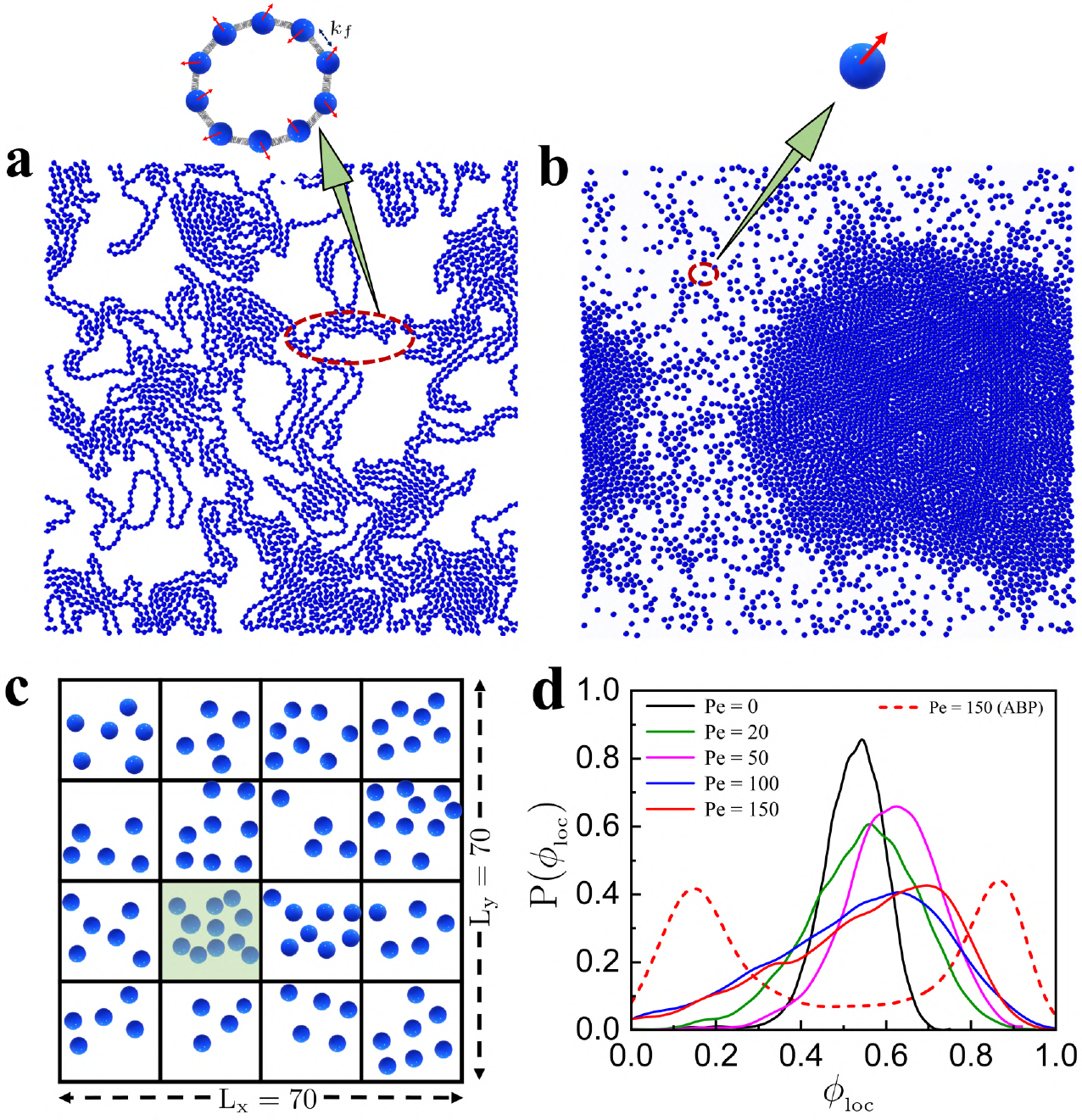}
\caption{\small Snapshots of a collection of (a) rings made of ABPs and (b) ABPs for $\text{Pe} = 150$ at $\phi = 0.48$. The red arrows represent the instantaneous directions of active force. A schematic sketch of a ring and an ABP is shown. The snapshots display the emergent dynamic clustering of active rings and motility-induced phase separation of ABPs for $\text{Pe} = 150$ at the same $\phi = 0.48$. (c) Schematic showing the method used for local density calculation, where $\phi_{_\text{loc}}$ is the number of particles in the shaded region divided by the area of that region. (d)$\text{P}(\phi_{_\text{loc}})$ $vs$ $\phi_{_\text{loc}}$ for different Pe at $\phi = 0.48$. The dashed line represents the curve for ABPs of the same $\phi$.}\label{fig:schematic_locD_pe}
\end{figure}

\noindent We model the rings as a sequence of $n$ number of ABPs of diameter $\sigma$ at positions $r_i$ $(i = 1, 2,.., n)$ that are connected by $n$ finitely extensible springs (Fig.~\ref{fig:schematic_locD_pe}a(i)). In contrast to existing efforts of simulating active dynamics of ring polymers, our model does not impose any external polarity~\cite{wen2023collective} or tangential activity~\cite{bianco2018globulelike}. Rather, it captures the emergent properties of the ensemble of active rings with random motility and therefore is quite unique. We pack $N$ number of such rings in a 2D square box of length $L_x = L_y = 70\sigma$ to achieve different packing fractions, $\phi$. The motion of each particle (monomer) is governed by the overdamped Langevin equation: 
\begin{equation}
\gamma \frac{d \textbf{r}_{i}}{dt} = - \sum_{j} \nabla V(\textbf{r}_i-\textbf{r}_j) + {\bf f}_{i}(t) + {\bf{F}_{\text{a, i}}(t)}
\label{eq:langevineq}
\end{equation}
where the drag force, $\gamma \frac{d \textbf{r}_{i}}{dt}$ is the velocity of $i^{th}$ monomer times the friction coefficient $\gamma$, and the total interaction potential $V(r) = V_{\text{FENE}} + V_{\text{BEND}} + V_{\text{WCA}}$ consists of bond, bending, and excluded volume contributions (see SM for details). Thermal fluctuations are captured by the Gaussian random force $f_i(t)$, which must satisfy the fluctuation-dissipation theorem. The activity is modeled as a propulsive force ${\text{F}_{\textrm{a}} \bf n(\boldsymbol \theta_{i})}$ on each monomer where $\text{F}_{\text{a}}$ represents the amplitude of active force with orientation specified by the unit vector $\bf n(\boldsymbol \theta_{i})$ evolves according to thermal rotational diffusion~\cite{bechinger2016active} (see SM for details). In our simulation, $\sigma$, $k_B T$, and $\tau=\frac{\sigma^2 \gamma}{k_B T}$ set the unit of length, energy, and time scales, respectively. We express the activity in terms of a dimensionless quantity, P\`{e}clet number as $\text{Pe} = \frac{\text{F}_a \sigma}{k_B T}$~\cite{theeyancheri2023active,theeyancheri2022migration}.

\section*{R\lowercase{esults and} D\lowercase{iscussion}}

\noindent We first simulate ABPs in two dimensions for different activities for a given packing fraction, $\phi = 0.48$. Where, $\phi = \frac{N n \pi \sigma^2}{4 L_x \times L_y}$, where $N$ is the number of rings and $n$ is the number of monomers with diameter $\sigma$ per rings. Our results show that ABPs undergo nonequilibrium clustering similar to other model active systems. We establish that this clustering is indeed activity-driven phase separation by estimating the local density, $\phi_{_\text{loc}}$ for different $\text{Pe}$ (Fig.~\ref{fig:schematic_locD_pe}d). For lower activities, the system behaves like a fluid, and $\text{P}(\phi_{_\text{loc}})$ has a single peak around the bulk density, but for higher activities, ABPs start to form clusters and the local density distribution changes from unimodal to bimodal~\cite{redner2013structure}. This signifies the phase separation and existence of two distinct phases in the system.\\ 
\begin{figure}[h!]
\centering
\includegraphics[width=0.99\linewidth]{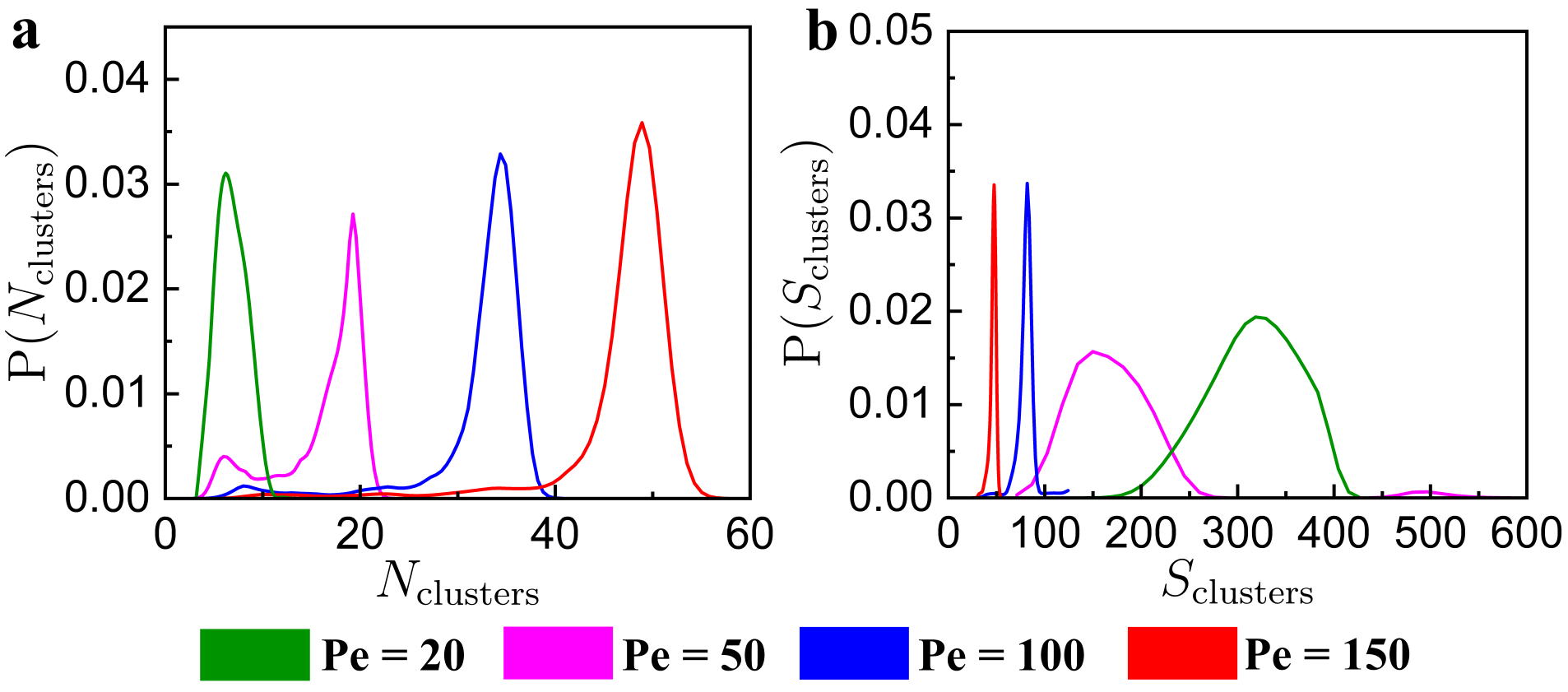}
\caption{\small (a) $\text{P}(N_\text{clusters})$ $vs$ $N_\text{clusters}$ and (b) $\text{P}(S_\text{clusters})$ $vs$ $S_\text{clusters}$ for different Pe at $\phi = 0.48$.}\label{fig:clusters_pe}
\end{figure}
\begin{figure*}
\centering
\includegraphics[width=0.99\linewidth]{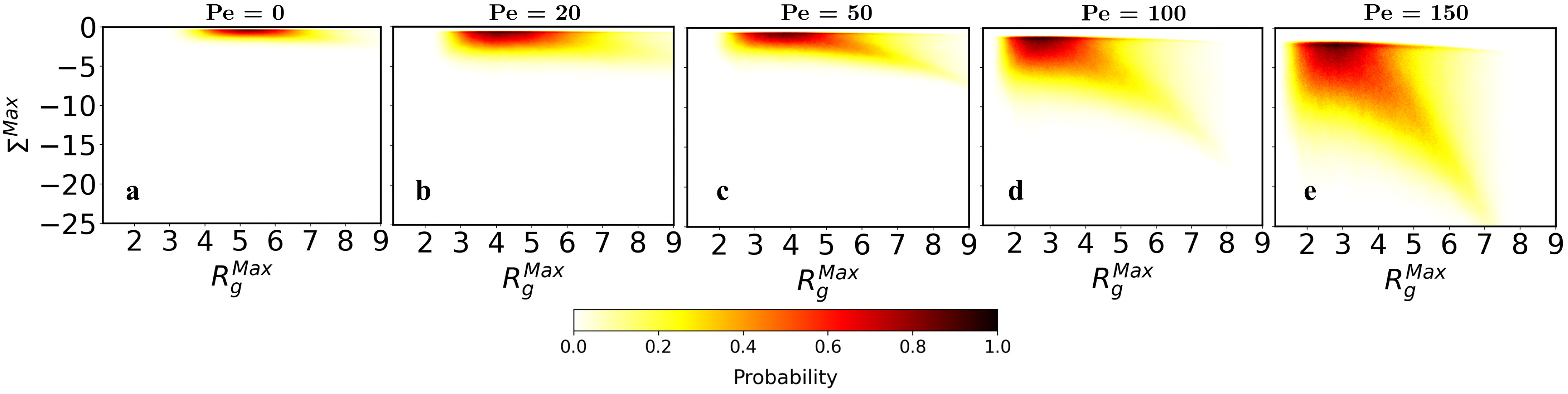}
\caption{\small 2D probability of $R_g^{\text{Max}}$ and that of $\Sigma^{\text{Max}}$ for (a) $\text{Pe = 0}$, (b) $\text{Pe = 20}$, (c) $\text{Pe = 50}$, (d) $\text{Pe = 100}$, and (e) $\text{Pe = 150}$ at $\phi = 0.48$.}\label{fig:stress_pe}
\end{figure*}

\noindent Next, we investigate the behavior of an ensemble of active Brownian rings at the same $\phi$ as ABPs to see how the conventional MIPS-like behavior is influenced by connecting the ABPs together to form the rings. For the passive rings, the peak of the $\text{P}(\phi_{_\text{loc}})$ occurs at $\phi_{_\text{loc}}=0.48$ i.e., at the bulk density $\phi$.
In the presence of activity, $\text{P}(\phi_{_\text{loc}})$ broadens with peak shifted to higher density at very high $\text{Pe}$ (Fig.~\ref{fig:schematic_locD_pe}d). Even for $\text{Pe} = 150$, where ABPs show MIPS, active rings do not exhibit two distinct phases. This denotes a dynamic clustering of active rings as a function of increasing $\text{Pe}$. Active rings form low and highly dense states (Fig.~S2), and these states are not stable as the cluster degradation happens, which is more pronounced with increasing $\text{Pe}$ (Movie\_S1-S2). However, the center of mass of the rings shows enhanced dynamics with intermediate superdiffusion as a function of increasing $\text{Pe}$ (Fig.~S3). \\ 

\noindent To further characterize the nature of the dynamic clustering of active rings, we compute the probability distribution of the number of clusters $\text{N}_\text{clusters}$, $\text{P}(\text{N}_\text{clusters})$ and size of the clusters $\text{S}_\text{clusters}$, $\text{P}(\text{S}_\text{clusters})$ formed for different $\text{Pe}$. If a minimum of 4 particles of the same or different rings are closely spaced within a cutoff distance of 1.12$\sigma$, then we defined them as a cluster. We observe a broader density profile, and the peak shifts towards the larger value for $\text{P}(\text{N}_\text{clusters})$ with increasing $\text{Pe}$, while a reverse trend can be seen in $\text{P}(\text{S}_\text{clusters})$, where the distribution become narrower and peaks shift to lower values of $\text{S}_\text{clusters}$ (Fig.~\ref{fig:clusters_pe}). This indicates the emergence of multiple dynamic clusters, and they continuously coalesce, split, and decay but never merge and grow into a stable dense phase like ABPs.\\

\noindent To elucidate the dynamic clustering, we consider the conformations and shape deformation of the active rings for different $\text{Pe}$ (Fig.~S4). We generate the probability distribution of the radius of gyration $R_g$, $\text{P}(R_g)$. $\text{P}(R_g)$ demonstrates the activity-induced collapse of the rings as the most probable $R_g$ shifts to smaller values with increasing $\text{Pe}$ (Fig.~S4a). We also calculate the asphericity parameter, $\text{A} = \frac{(\lambda_2 - \lambda_1)^2}{(\lambda_1 + \lambda_2)^2}$, where $\lambda_1 \, \text{and} \, \lambda_2$ are the eigenvalues of the gyration tensor. $\text{P(A)}$ indicates that the extent of the ring deformation from the circular one also increases with the increase in $\text{Pe}$. Active rings undergo moderate deformation at lower $\text{Pe}$ followed by more stronger deformation within a small fraction of rings at high $\text{Pe}$ (Fig.~S4b). The rings are coming closer in the cluster due to activity which leads to the shape deformation and the collapse of the rings. \\

\noindent Subsequently, we examine how deformation and stress are distributed within the system in Fig.~\ref{fig:stress_pe}, where we plot the maximum eigenvalues of stress tensor ($\Sigma^{\text{Max}}$) against maximum eigenvalues of gyration tensor ($R_g^{\text{Max}}$).  Fig.~\ref{fig:stress_pe} displays that the eigenvalues of the stress and the gyration tensor are correlated so that the stress increases to a large extent, rings deform and shrink with increasing $\text{Pe}$. The negative values of the stress indicate that this is of a compressive nature. For passive rings, the stress contribution is from the bonded interactions as there is no clustering (Fig.~\ref{fig:stress_pe}a). On the other hand, active rings display an enhancement in stress due to the deformation-induced shrinking with increasing $\text{Pe}$.\\
\begin{figure}
\centering
\includegraphics[width=0.99\linewidth]{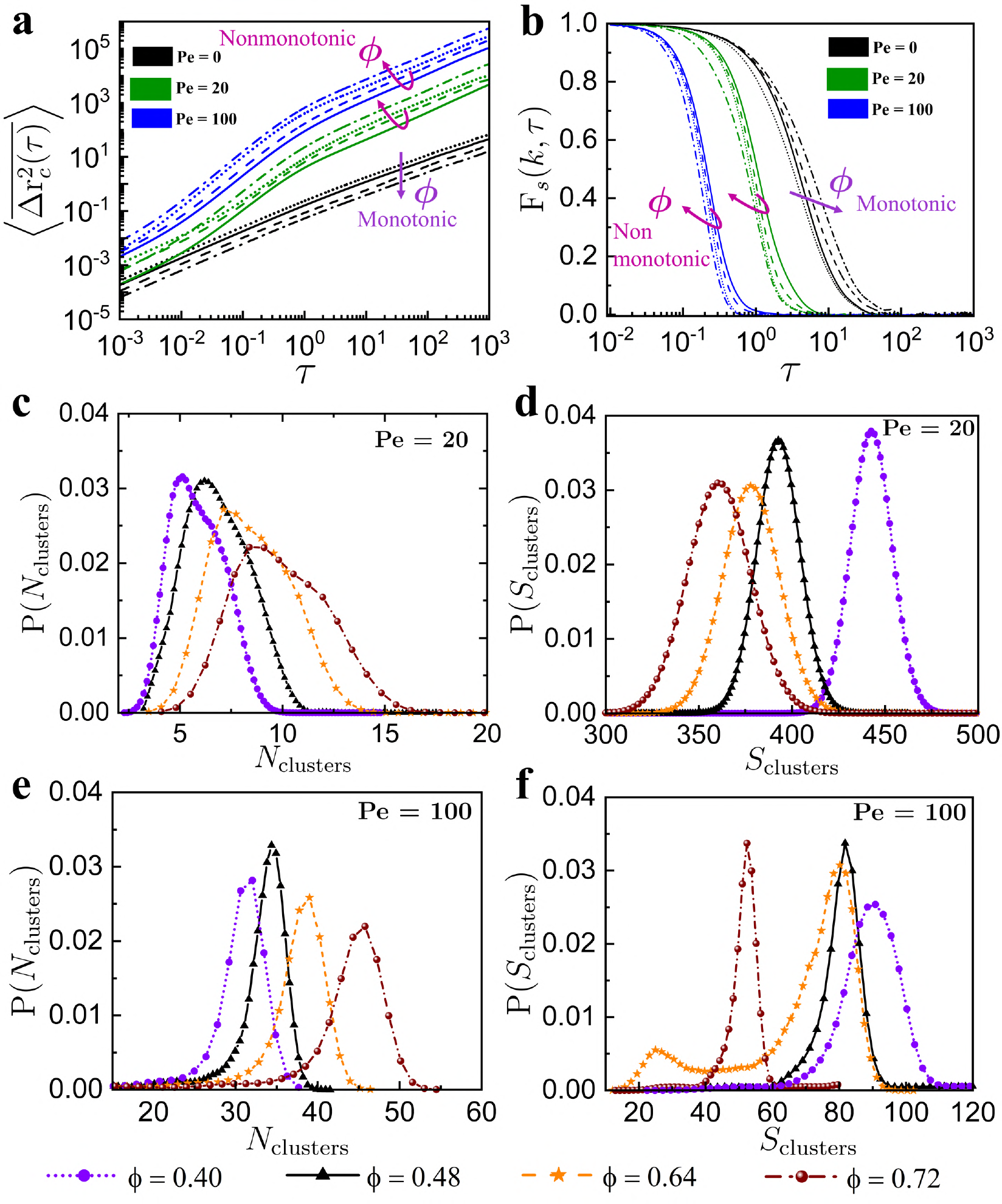}
\caption{\small (a) Log-log plot of $\left\langle{\overline{\Delta r_{c}^{2}(\tau)}}\right\rangle$, (b) log-linear plot of $\text{F}_s(k, \tau)$ $vs$ $\tau$ for $\text{Pe = 0}$, $\text{Pe = 20}$, and $\text{Pe = 100}$ for $\phi = 0.40$ (dotted), $\phi = 0.48$ (solid), $\phi = 0.64$ (dashed), and $\phi = 0.72$ (dash-dotted). The magenta and violet colored arrows indicate the nonmonotonic and monotonic behavior in the dynamics of active rings as a function of increasing $\phi$. $\text{P}(N_\text{clusters})$ $vs$ $N_\text{clusters}$ and $\text{P}(S_\text{clusters})$ $vs$ $S_\text{clusters}$ for $\text{Pe = 20}$ (c and d), and $\text{Pe = 100}$ (e and f) for different $\phi$.}\label{fig:dyn_cluster_phi}
\end{figure}

\noindent The results derived so far demonstrate how the motion of a collection of active rings deviates from the conventional behavior exhibited by ABPs as a function of $\text{Pe}$. Now, we examine how the packing fraction ($\phi$) influences the dynamics and clustering of active rings. We compute the center of mass mean square displacement, $\left\langle{\overline{\Delta r_{c}^{2}(\tau)}}\right\rangle$ of the rings with increasing $\phi$ (Fig.~\ref{fig:dyn_cluster_phi}a). For the passive rings, the dynamics becomes slower with increasing $\phi$ and consequently, $\left\langle{\overline{\Delta r_{c}^{2}(\tau)}}\right\rangle$ show a monotonic decrease with $\phi$. Conversely, we notice a nonmonotonic behavior in the dynamics of active rings where $\left\langle{\overline{\Delta r_{c}^{2}(\tau)}}\right\rangle$ first decreases up to an intermediate $\phi$ and then increases upon further increasing $\phi$. The non-monotonic behavior observed for active rings is directly correlated to the breakdown of the clusters in smaller ones at large $\phi$. The cluster number and size for $\text{Pe = 20}$ (Fig.~\ref{fig:dyn_cluster_phi}(c, d)) and $\text{Pe = 100}$ (Fig.~\ref{fig:dyn_cluster_phi}(e, f)) display formation of a large number of smaller clusters with increasing $\phi$ (Movie\_S2-S4). The peaks of $\text{P}(N_\text{clusters})$ shift to higher values of $N_\text{clusters}$, while the most probable values of $\text{P}(S_\text{clusters})$ shift to smaller values of $S_\text{clusters}$ as function of increasing $\phi$. This behavior is more pronounced at higher $\text{Pe}$ as the cluster formed are much smaller in size compared to lower $\text{Pe}$ for all $\phi$. This signifies the formation of highly motile smaller and dense clusters at higher $\phi$ (Fig.~S5), which in turn results in faster motion of the active rings. We also report the self-intermediate scattering functions $\text{F}_s(k, \tau)$ for $k$ corresponds to the first peak of the radial distribution functions for different $\phi$. $\text{F}_s(k, \tau)$ decays slowly followed by a fast decay with increasing $\phi$ while the associated relaxation time $\tau_a$ first increases and then decreases with increasing $\phi$ (Fig.~S6), which further confirms the non-monotonic behavior observed for active rings (Fig.~\ref{fig:dyn_cluster_phi}b).

\subsection*{S\lowercase{ummary}}

\noindent Our in silico model identifies how the motion of an ensemble of active rings is regulated by the crucial structural and dynamic factors that control their motion. A dense suspension of ABPs separates into two distinct phases as seen in the local density plot (Fig.~\ref{fig:schematic_locD_pe}d). But the phenomenon is quite different when these ABPs are connected to form rings. This system does not show MIPS. Instead, an emergent dynamic clustering of rings is observed. The presence of springs prevents the system from undergoing MIPS. The interplay between activity (motility), topology, and crowding drives the system to form intermittent clusters but never allows grow into a large dense phase/cluster. Our simulations exhibit that the translational motion of the center of mass of the rings is enhanced by the activity for a given $\phi$. Interestingly, the behavior of the system changes drastically with increasing $\phi$. The translational dynamics of the system becomes faster at higher $\phi$ and also display an emergent dynamic clustering of active rings. The size and number of clusters largely depend on the magnitude of \text{Pe} and $\phi$. At higher $\phi$, dynamic clustering leads to the formation of highly motile smaller and dense clusters. Our model finds that the dynamic clustering is different from MIPS because the stress building in the dynamic clusters by the activity-induced deformation (Fig.~S7-S8) followed by the stress-releasing $via$ cluster degradation as the monomers are connected by springs, which suppresses the formation of stable dense and dilute regions of active rings. The deformation of rings and activity-assisted stress governing events lead to the emergence of distinctive dynamic clustering of active rings, which is completely absent in the passive case (Fig.~\ref{fig:stress_pe}a). We believe our study, in general will be insightful in understanding the structure and dynamics of densely packed deformable objects driven by motility.

\section*{A\lowercase{cknowledgments}}

\noindent L.T. thanks UGC for a fellowship and IIT Bombay for the institute postdoctoral fellowship. S.C. thanks DST Inspire for a fellowship. R.C. acknowledges SERB for funding (Project No. MTR/2020/000230 under MATRICS scheme). T.B. acknowledges NCBS-TIFR for research funding. L.T. thanks Pooja Nanavare for the help and proofreading. The authors thank Kiruthika Kumar for the initial discussions and for performing some trial simulations.\\

\section*{Data Availability}

\noindent The codes and data used for this paper are available from the authors upon request.


\appendix
\section*{Supplementary Material}
\renewcommand{\thefigure}{S\arabic{figure}}
\setcounter{figure}{0}
\noindent \hspace{18mm}\textbf{Model and Simulation Details} \\ 

\begin{figure}[h!] 
\centering 
\includegraphics[width=0.95\linewidth]{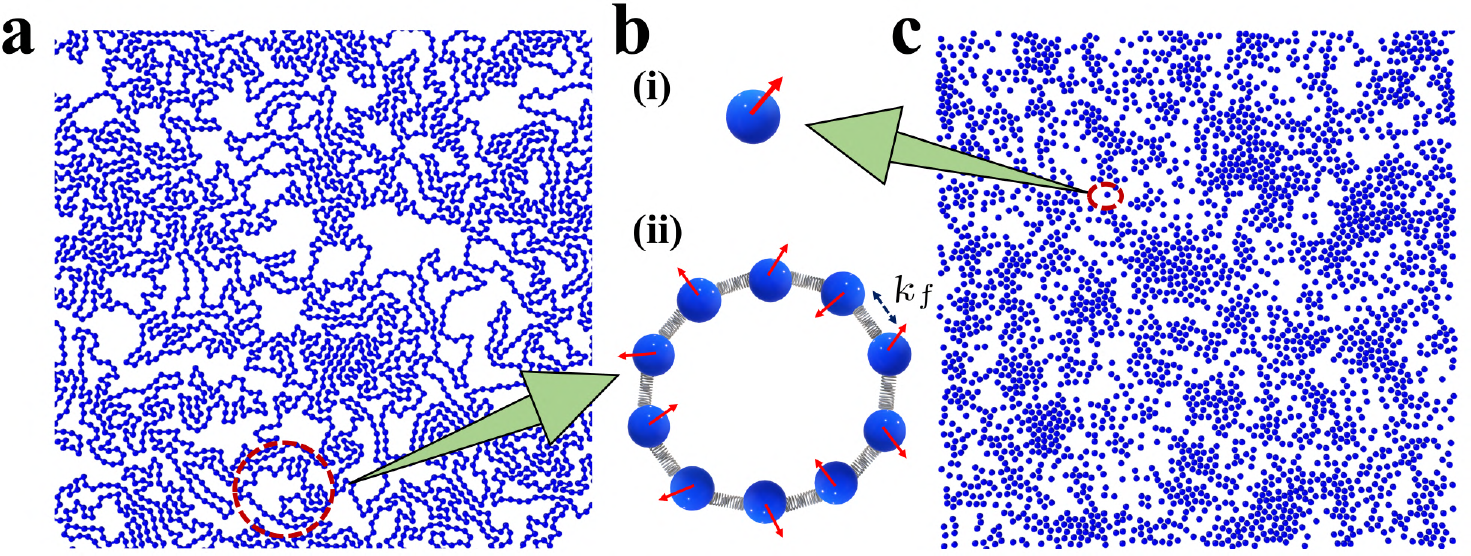} \\
\caption{\small A snapshot of a collection of (a)  ring polymers, (b) schematic sketch of an (i) ABP, (ii) an active ring, and (c) a collection of active Brownian particles. The red arrows represent the instantaneous directions of active force.}\label{fig:schematic_c5} 
\end{figure}

\noindent  We model the rings as a sequence of $n$ number of active Brownian particles of diameter $\sigma$ at positions $r_i$ $(i = 1, 2,.., n)$ that are connected by $n$ finitely extensible springs (Fig.~\ref{fig:schematic_c5}). We pack $N$ numbers of such rings in the 2D square box of length $L_x = L_y = 70\sigma$ to obtain a collection of rings with different packing fractions, $\phi$. The packing fraction is defined as $\phi = \frac{N n \pi \sigma^2}{4 L_x \times L_y}$, where $N$ is the number of rings and $n$ is the number of monomers with diameter $\sigma$ per rings. Similarly, we consider active Brownian particles (ABPs) of varying packing fractions ($\phi$) in the 2D square box. In our simulations, $\sigma$, $k_B T$ and $\tau=\frac{\sigma^2 \gamma}{k_B T}$ set the unit of length, energy, and time scales, respectively. Where $k_B$ is the Boltzmann constant, $T$ is the temperature, and $\gamma$ is the friction coefficient. The state of the system is represented by the positions $r_i$ of the active beads, and their evolution is governed by the following Langevin equation with an additional term to account for the activity. Here, we consider a high friction limit. Therefore, the dynamics is practically overdamped as the contribution from the inertia term is negligible, and hence we do not include the inertia term in the following equation of motion.
\begin{equation}
\gamma \frac{d \textbf{r}_{i}}{dt} = - \sum_{j} \nabla V(\textbf{r}_i-\textbf{r}_j) + {\bf f}_{i}(t) + {\bf{F}_{\text{a, i}}(t)}
\label{eq:langevineq_c5}
\end{equation}
where the drag force, $\gamma \frac{d \textbf{r}_{i}}{dt}$ is the velocity of each bead times the friction coefficient $\gamma$, $V(r)$ is the resultant pair potential between $i^\text{th}$ and $j^\text{th}$ particles, thermal force $\bf f_{i}(t)$ is modeled as Gaussian white noise with zero mean and variance $\left<f_{i}(t^{\prime})f_{j}(t^{\prime\prime})\right> = 4 \gamma k_B T \delta_{ij}\delta(t^{\prime}-t^{\prime\prime})$, and ${\text{F}_{\text{a, i}}(t)}$ is the active force which drives the system out of equilibrium. ${\bf{F}_{\text{a, i}}(t)}$ has the magnitude $\text{F}_{\text{a}}$, acts along the unit vector~\cite{bechinger2016active} of each $i^{th}$ monomer, $\bf{n}(\boldsymbol{\theta_i}) = {(\textrm{cos} \,\boldsymbol{\theta_i}, \, \textrm{sin} \, \boldsymbol{\theta_i})}$, where $\theta_i$ evolves as $\frac{d \boldsymbol{\theta_i}}{dt} = \sqrt{2D_R} {\bf f}_{i}^{R}$, $D_R$ is the rotational diffusion coefficient and ${\bf f}_{i}^{R}$ is the Gaussian random number with a zero mean and unit variance. Therefore, the persistence time, $\tau_R$ of the individual monomers is related to $D_R$ as $\tau_R = \frac{1}{D_R}$. The activity can also be expressed in terms of a dimensionless quantity, i.e., the P\'eclet number Pe, which is defined as $\frac{\text{F}_a \sigma}{k_BT}$. The total interaction potential $V(r) = V_{\text{FENE}} + V_{\text{BEND}} + V_{\text{WCA}}$ consists of bond, bending, and excluded volume contributions. Nearest-neighbour monomers along the contour of the rings are connected by the FENE potential:
\begin{equation}
V_{\text{FENE}}\left(r_{ij}\right)=\begin{cases} -\frac{k_f r_{\text{max}}^2}{2} \ln\left[1-\left( {\frac{r_{ij}}{r_{\text{max}}}}\right) ^2 \right],\hspace{3mm} \mbox{if } r_{ij} \leq r_{\text{max}}\\
\infty, \hspace{35mm} \mbox{otherwise}.
\end{cases}
\label{eq:FENE_c5}
\end{equation}
where $k_f$ is the spring constant, and $r_{ij}$ is the distance between two neighboring monomers in the ring with a maximum extension of $r_{\text{max}} = 1.5 \sigma$~\cite{kremer1990dynamics}. To account for self-avoidance a pair of monomers of the rings interact $via$ the repulsive Weeks–Chandler–Andersen (WCA) potential~\cite{weeks1971role}.
\begin{equation}
V_{\textrm{WCA}}(r_{ij})=\begin{cases}4\epsilon_{ij} \left[\left(\frac{\sigma_{ij}}{r_{ij}}\right)^{12}-\left(\frac{\sigma_{ij}}{r_{ij}}\right)^{6}\right]+\epsilon, \mbox{if }r_{ij}<2^{1/6}\sigma_{ij} \\
0, \hspace{35mm} \mbox{otherwise},
\end{cases}
\label{eq:WCA_c5}
\end{equation}
where $r_{ij}$ is the separation between the interacting particles, $\epsilon_{ij} = 1$ is the strength of the steric repulsion, and $\sigma_{ij} = \frac{\sigma_i + \sigma_j}{2}$ determines the effective interaction diameter, with $\sigma_{i(j)}$ being the diameter of the interacting pairs. We consider the size of the ring as $n = 50$, and vary the packing fraction of the rings from $\phi = 0.40, 0.48, 0.64, \, \text{and} \, 0.72$.\\

\noindent The equilibrium and kinetic properties of these systems are studied using fixed-volume and constant-temperature molecular dynamics simulations. The system dynamics is integrated by the velocity Verlet algorithm in each time step using LAMMPS~\cite{plimpton1995fast} engine with Langevin thermostat. We initialize the system by randomly placing $N$ number of rings in the square box and relaxing the initial configuration for $2 \times 10^6$ steps. All the production simulations are carried out for $10^9$ steps where the elementary integration time step is chosen to be $10^{-5}$, and the positions of the particles are recorded every $100$ step.\\

\noindent \hspace{10mm}\textbf{Self-intermediate Scattering Function}\\

\noindent We compute the self-intermediate scattering function $\text{F}_s(k, \tau)$ for the center of mass of the rings defined as. 
\begin{equation}
\text{F}_s(k, \tau)=\left<\frac{1}{N}\sum_{i=1}^{N}e^{{ik}.(r_c^i(t+\tau)-r_c^i(\tau))}\right>
\end{equation}
\noindent where N is the number of rings, $r_c$ is the center of mass of the ring, and for computation, we chose k values corresponding to the first peaks of respective radial distribution functions. $\text{F}_s(k, \tau)$ provides the characteristic relaxation time of the rings. It is computed for individual rings and averaged over all the rings in the system for different time frames. The associated relaxation time $\tau_a$ provides the details of how the system relaxes for different parameters. \\

\noindent \hspace{18mm} \textbf{Gyration Tensor Analysis} \\

\noindent To evaluate the shape fluctuations or how the rings deform, we calculate the gyration tensor of the conformations defined as,
 \begin{equation}
S = \left (\begin{smallmatrix}
 \sum_i (x_i-x_\text{com})^2 & \sum_i (x_i-x_\text{com})(y_i-y_\text{com}) \vspace{4mm} \\ \sum_i (x_i-x_\text{com})(y_i-y_\text{com}) & \sum_i (y_i-y_\text{com})^2 
\end{smallmatrix}\right )
\end{equation}

\noindent where, $x_\text{com} \, \text{and} \, y_\text{com}$ represent the x and y components of the COM position respectively. Further, we compute the eigenvalues $\lambda_1 \, \text{and} \, \lambda_2$ of the gyration tensor by diagonalizing the matrix S. These eigenvalues are used to define the shape descriptor, asphericity parameter, $\text{A} = \frac{(\lambda_2 - \lambda_1)^2}{(\lambda_1 + \lambda_2)^2}$. It measures the deviation from the spherical symmetry. The asphericity parameter is 1 for a perfect rod-like or linear topology and 0 for a circular structure. \\

\noindent We extract the radius of gyration,
\begin{equation}
R_g = \left [ \frac{1}{N} \sum_{i = 1}^N (r_i - r_\text{com})^2\right ]^{\frac{1}{2}}
\end{equation}

\noindent from gyration tensor for each time-step of every simulation after the system reaches the steady state. Here $r_\text{com}$ is the center of mass of the rings. Then a single trajectory is created by stitching different individual trajectories together. This single trajectory is binned to construct a histogram from which the ensemble-averaged probability distribution $\text{P}(R_g)$ is obtained.\\

\noindent \hspace{18mm} \textbf{Stress Tensor Analysis} \\

\noindent The stress tensor in two dimensions provides the distribution of internal forces within a deformable body. The stress tensor matrix can be defined as,
 \begin{equation}
\Sigma = \left (\begin{smallmatrix}
 \Sigma_{xx} & \Sigma_{xy} \vspace{5mm} \\ \Sigma_{yx} & \Sigma_{yy} 
\end{smallmatrix}\right )
\end{equation}

\noindent Here, the components represent the stresses acting on the planes perpendicular to the x and y axes. The subscripts indicate the direction of the stress components. $\Sigma_{xx}$ and $\Sigma_{yy}$ is the normal stresses in the x and y directions, $\Sigma_{xy}$ and $\Sigma_{yx}$ represent the shear stress acting on the plane with normal in the x(y) direction and parallel to the y(x) direction, respectively. Stress components are computed by taking in to account the forces experienced by the particle. Thus, 
\begin{equation}
\Sigma_{xx} = \frac{\textbf{F}_x}{A}   \hspace{15mm} \Sigma_{yy} = \frac{\textbf{F}_y}{A}
\end{equation}

\noindent here $\textbf{F}_x$ ($\textbf{F}_y$) is the total force experienced by all the monomers of the ring in the x(y) direction, and $A$ is the cross-sectional area perpendicular to the force. \\

\noindent To monitor the single ring stress behavior, we calculate the associated stress tensor where, for each ring, the monomer-monomer and non-bonded interaction forces are accounted. We calculated the stress tensor for individual rings and average over all the rings in the system. The resulting normalized eigen values are utilized for predicting the deformation of the rings in the densely packed systems. We show that the eigenvalues of the stress and of the gyration tensors are correlated, which implies that the deformation and stress are directly linked. Hence, a larger stress corresponds to a larger deformation (and asymmetry). The negative eigenvalues of the stress indicate that this is of compressive nature. \\

\begin{figure*}
\centering
\includegraphics[width=0.95\linewidth]{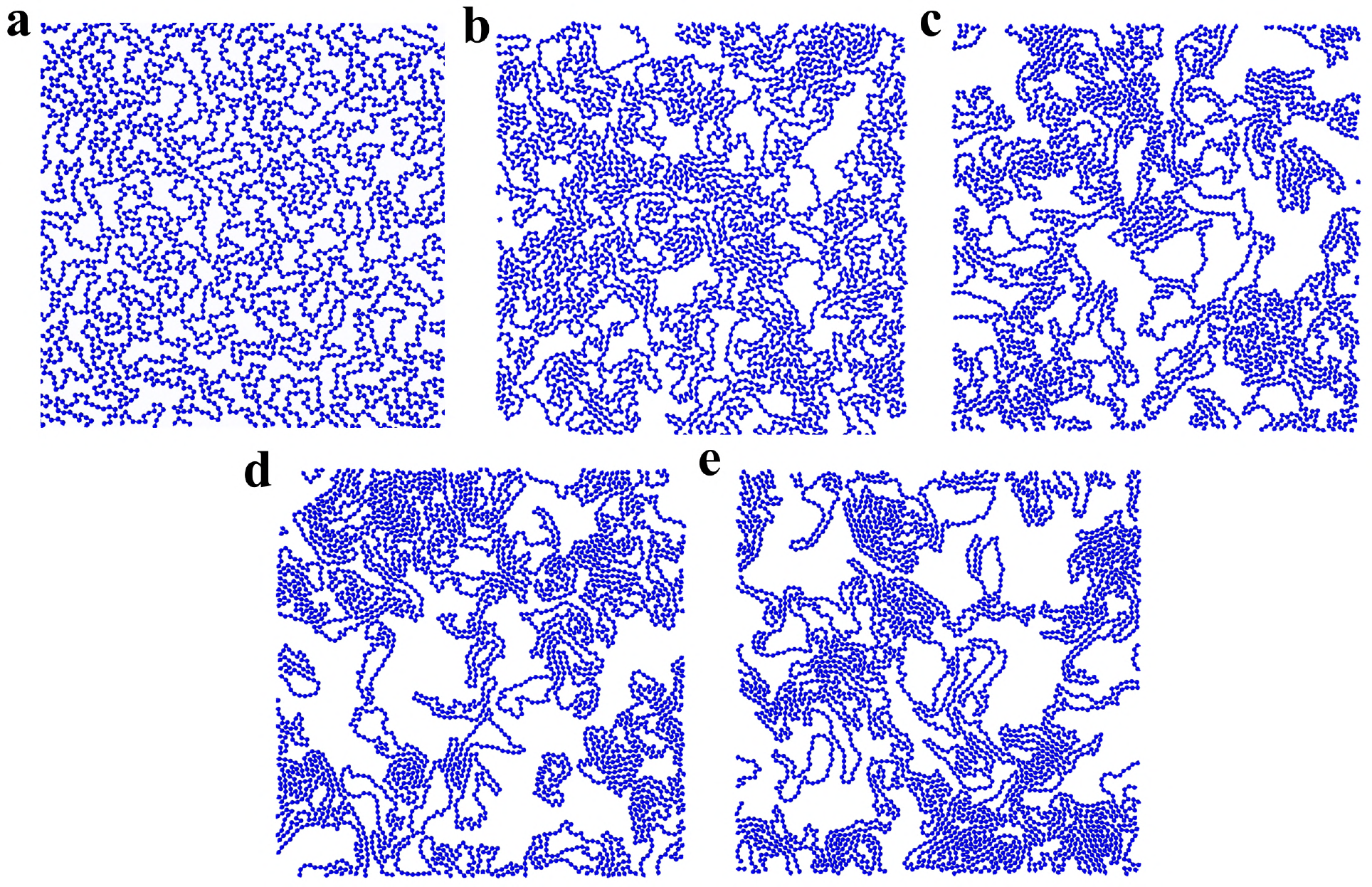}
\caption{\small Snapshots of active rings for (a) $\text{Pe = 0}$, (b) $\text{Pe = 20}$, (c) $\text{Pe = 50}$, (d) $\text{Pe = 100}$, and (e) $\text{Pe = 150}$ at $\phi = 0.48$.}\label{fig:snapshot_pe}
\end{figure*}
\begin{figure*} 
\centering
\includegraphics[width=0.95\linewidth]{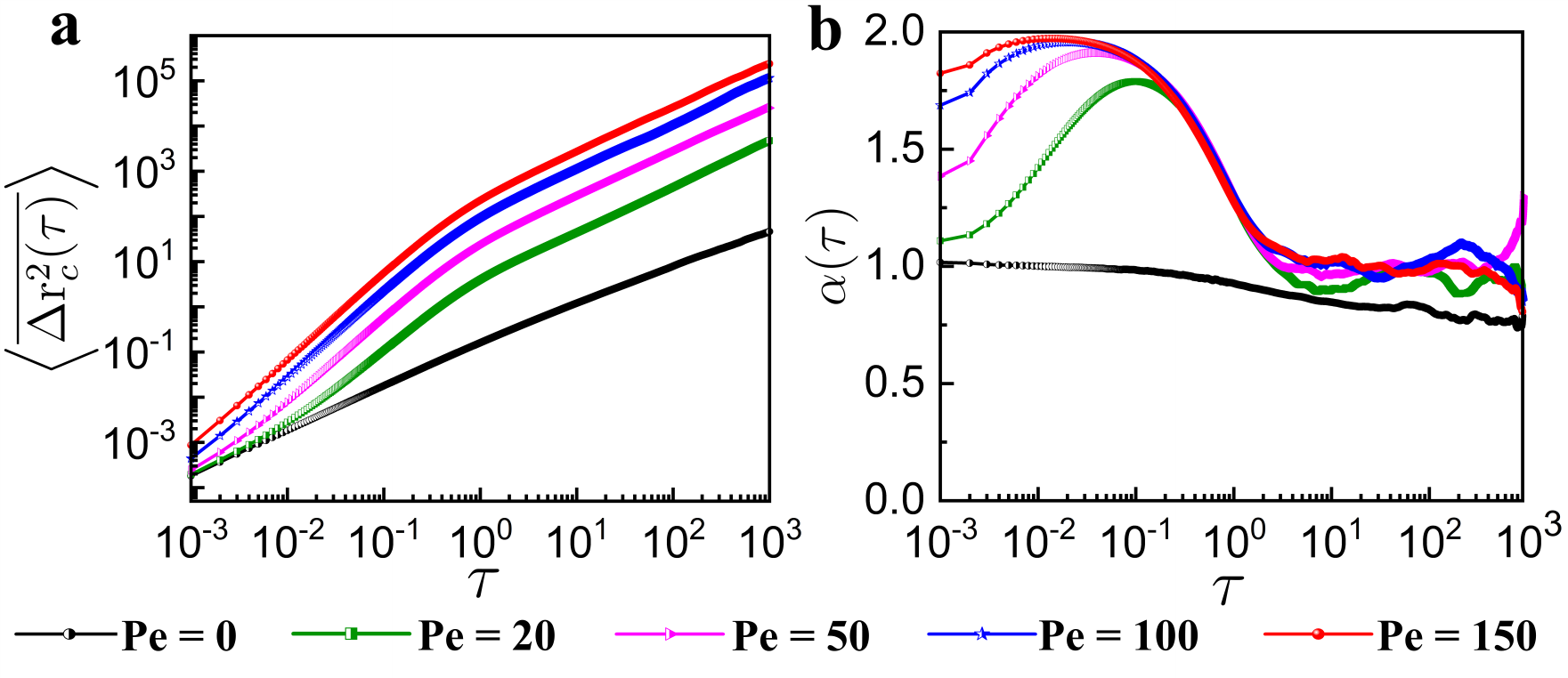}
\caption{\small (a) Log-log plot of $\left\langle{\overline{\Delta \text{r}_{c}^{2}(\tau)}}\right\rangle$ and (b) $\alpha(\tau)$ $vs$ $\tau$ of the rings subjected to different $\text{Pe}$ for $\phi = 0.48$.}\label{fig:dyn_pe}
\end{figure*}
\begin{figure*}
\centering 
\includegraphics[width=0.88\linewidth]{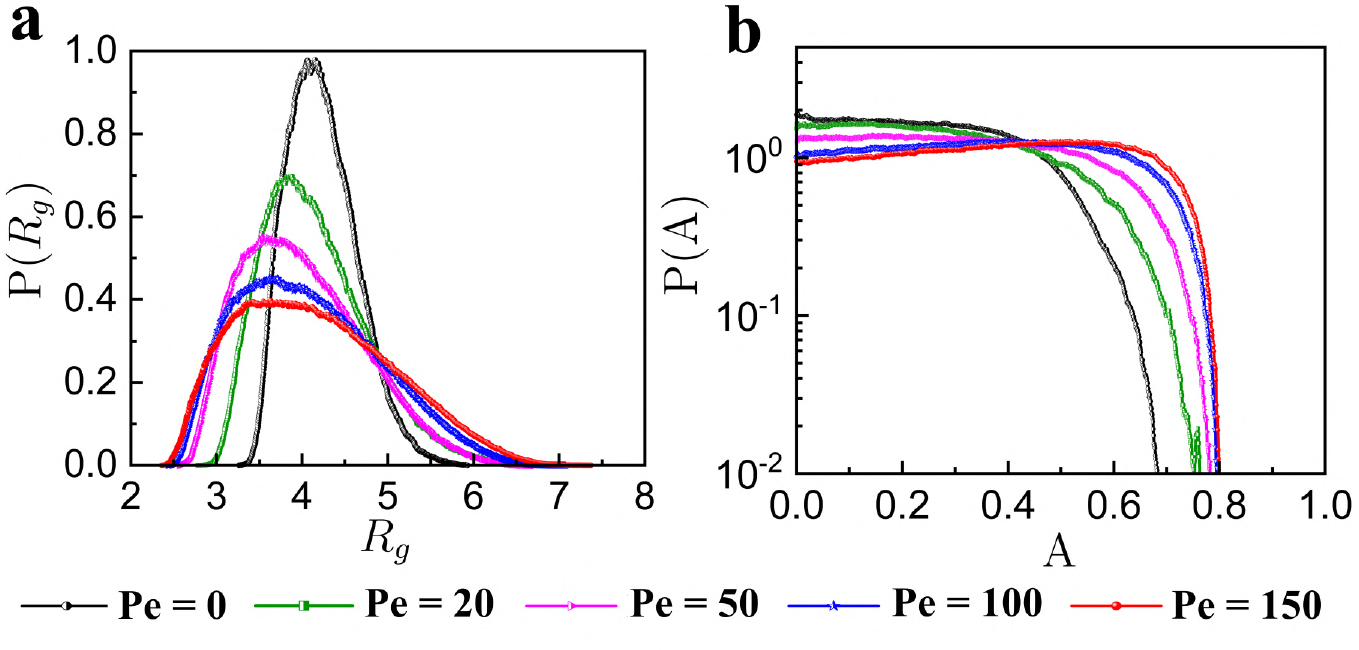}
\caption{\small (a) $\text{P}(\text{R}_g)$ $vs$ $\text{R}_g$ and (b) $\text{P}(\text{A})$ $vs$ $\text{A}$ of rings subjected to different \text{Pe}.}\label{fig:dist_rg_pe}
\end{figure*}
\begin{figure*}
\centering
\includegraphics[width=0.9\linewidth]{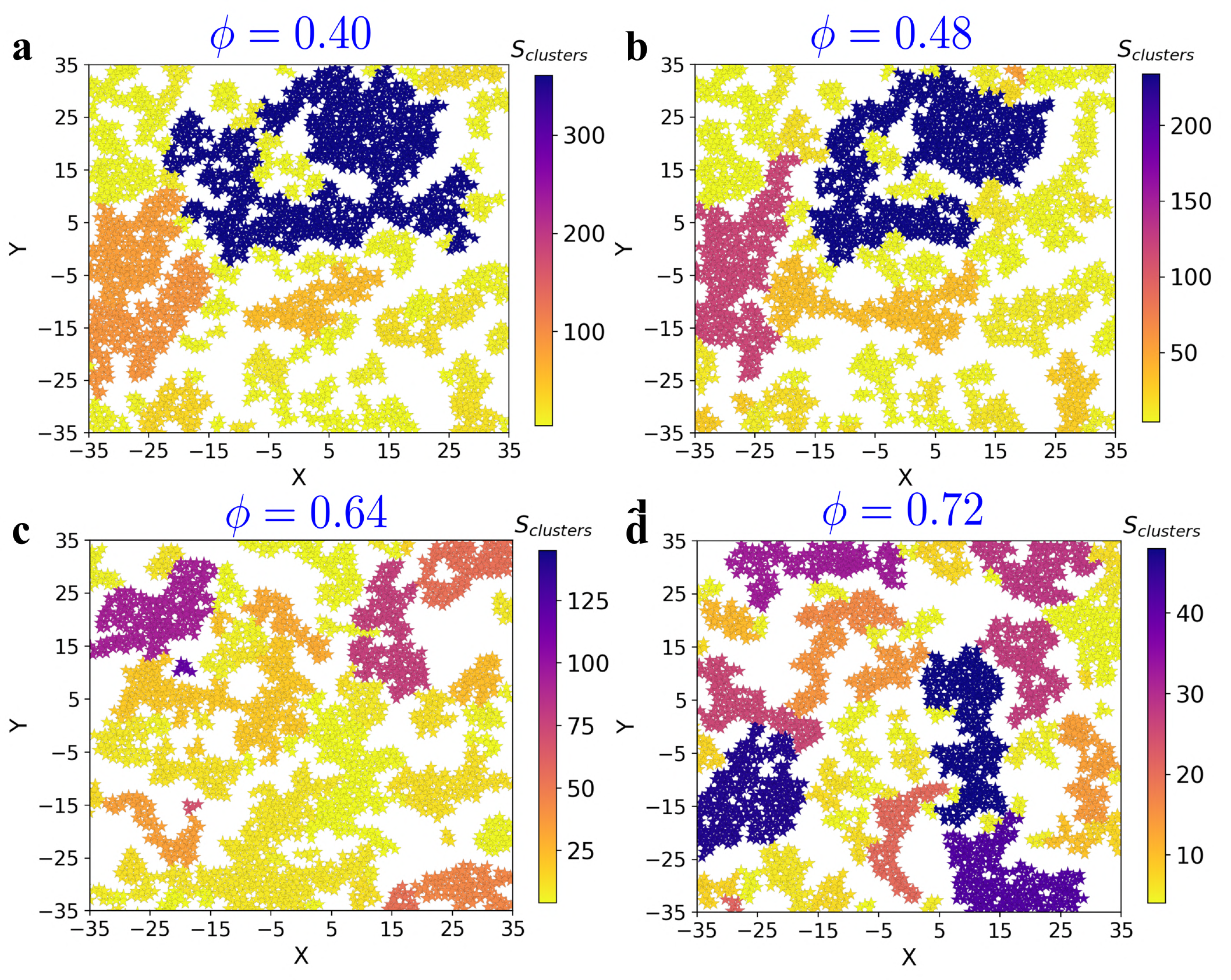}
\caption{\small Color map showing the $\text{S}_\text{clusters}$ of active ($\text{Pe = 100}$) rings for (a) $\phi = 0.40$, (b) $\phi = 0.48$, (c) $\phi = 0.64$, and (d) $\phi = 0.72$.}\label{fig:colormap_sc}
\end{figure*}
\begin{figure*}
\centering
\includegraphics[width=0.85\linewidth]{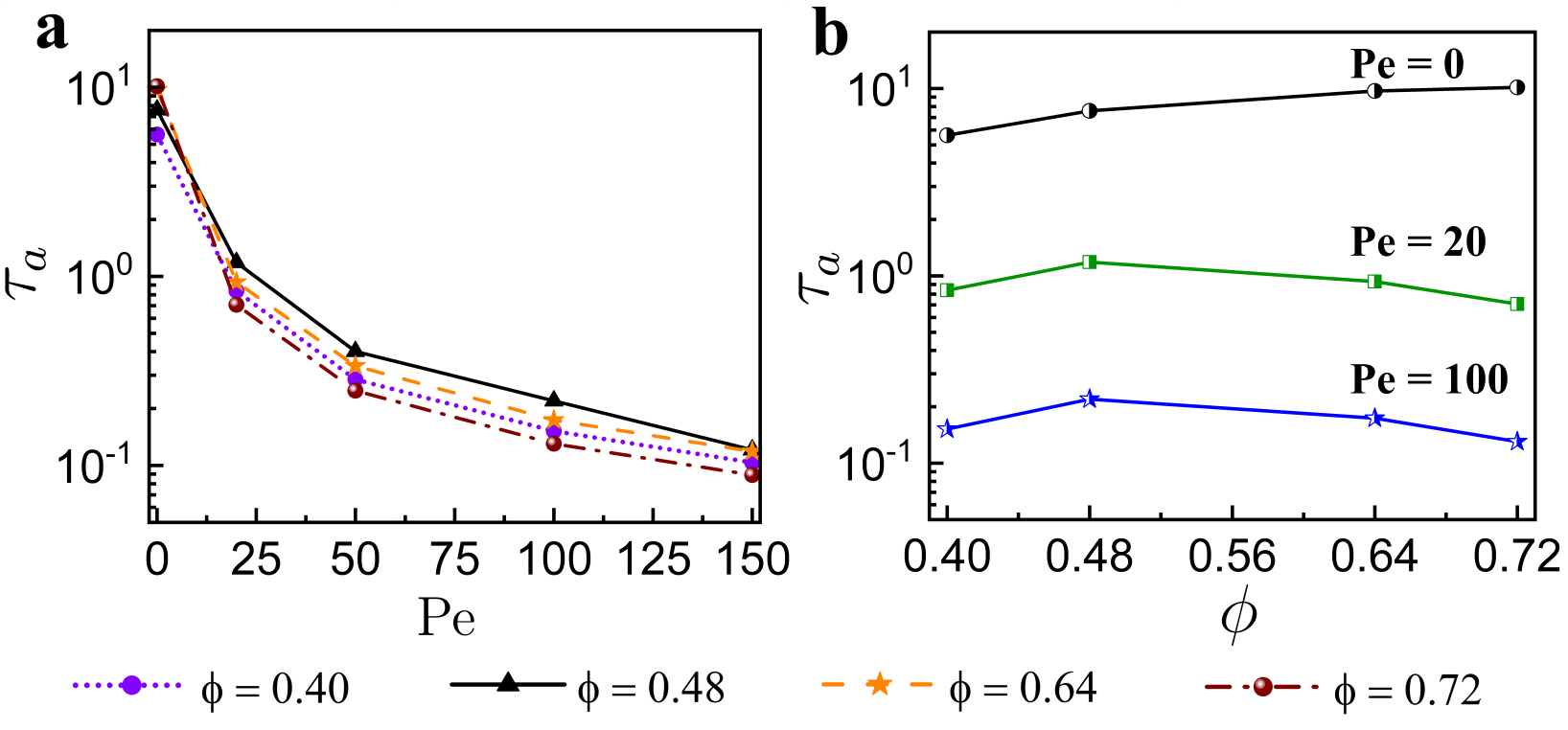}
\caption{\small The relaxation time $\tau_a$ of active rings (a) at different $\phi $ as function of increasing $\text{Pe}$ and (b) for $\text{Pe}$ = 0, 20 and 100 at different $\phi$.}\label{fig:tau_a_phi}
\end{figure*}
\begin{figure*}
\centering
\includegraphics[width=0.85\linewidth]{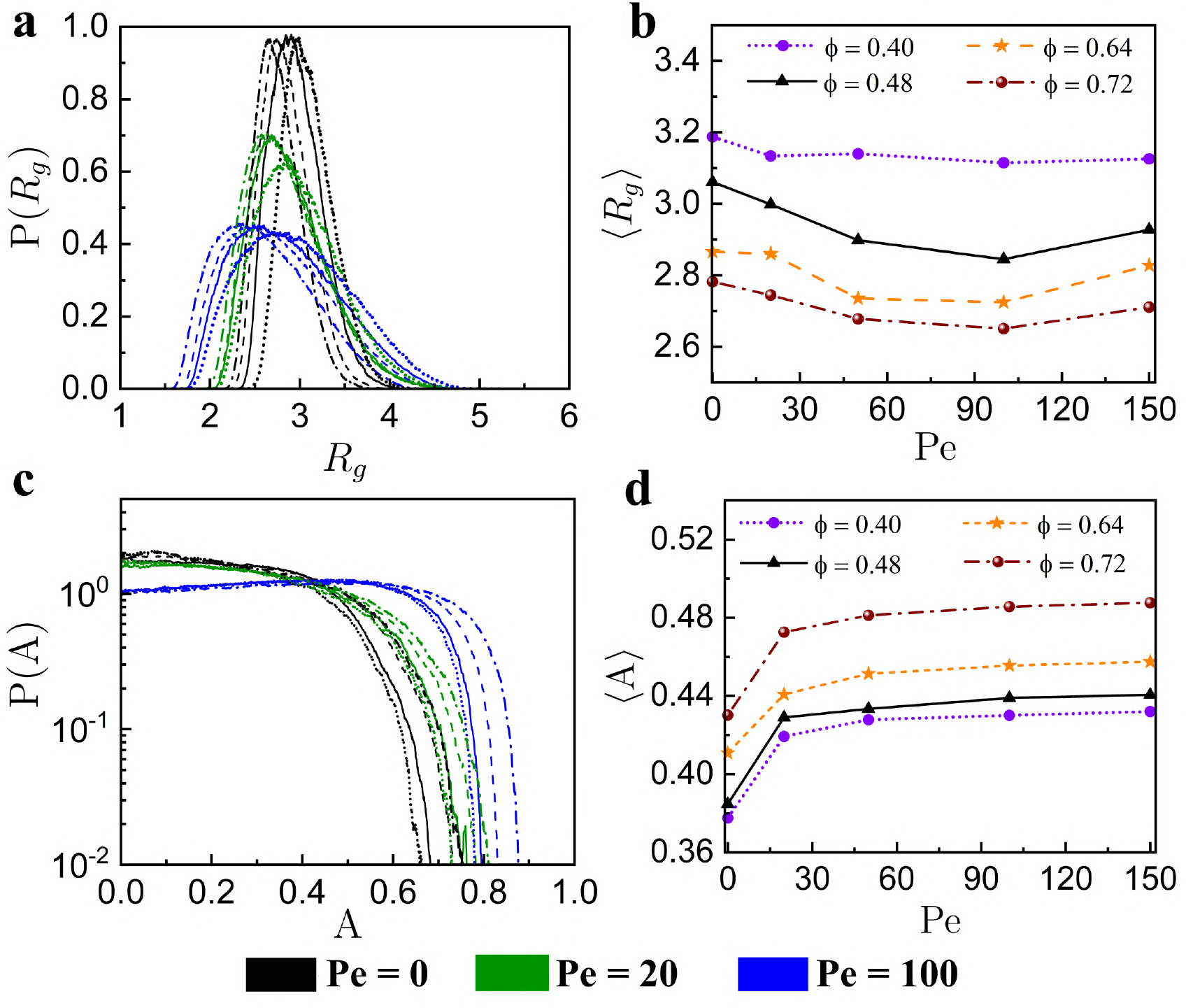}
\caption{\small (a) $\text{P}(\text{R}_g)$ $vs$ $\text{R}_g$, (b) $\left < \text{R}_g\right>$, (c) $\text{P}(\text{A})$ $vs$ $\text{A}$,  and (d) $\left < \text{A}\right>$ of rings subjected to different \text{Pe} for $\phi = 0.40$ (dotted), $\phi = 0.48$ (solid), $\phi = 0.64$ (dashed), and $\phi = 0.40$ (dash-dotted).}\label{fig:conf_phi}
\end{figure*}
\begin{figure*}
\centering
\includegraphics[width=0.95\linewidth]{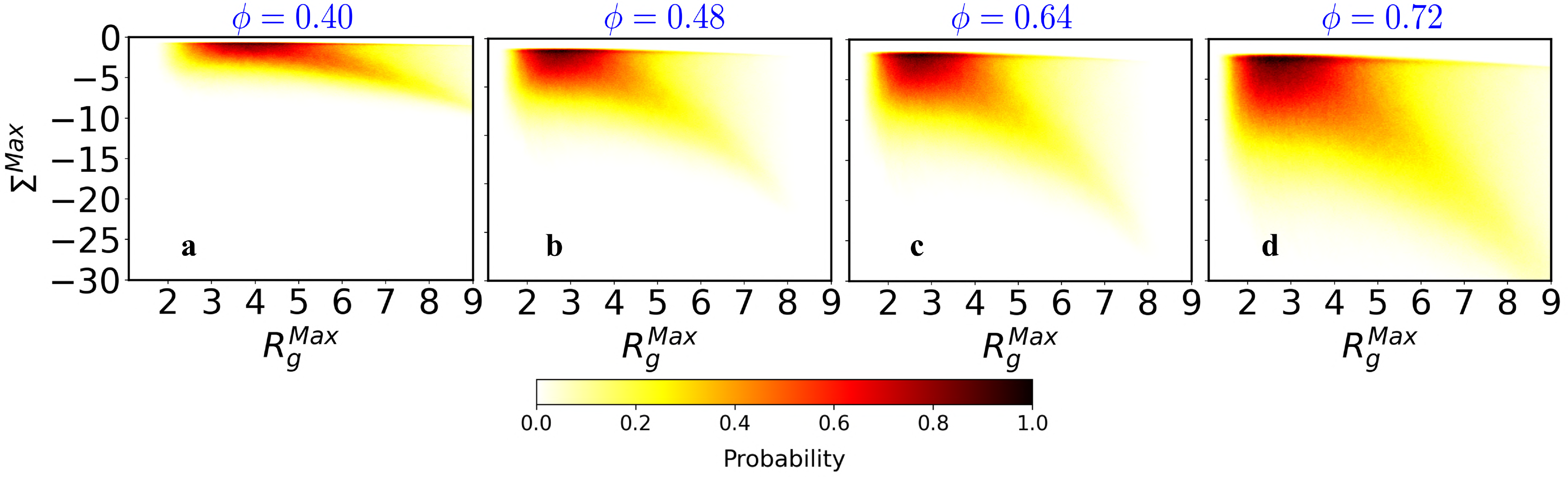}
\caption{\small 2D probability of $R_g^{\text{Max}}$ and that of $\Sigma^{\text{Max}}$ of active (\text{Pe = 100}) rings for (a) $\phi = 0.40$, (b) $\phi = 0.48$, (c) $\phi = 0.64$, and (d) $\phi = 0.72$.}\label{fig:stress_phi}
\end{figure*}

\noindent \hspace{28mm}\textbf{Movie Description} \\

\begin{enumerate}
\item \textbf{Movie\_S1} \\
\noindent The motion of passive ($\text{Pe = 0}$) rings ($n = 50, \, N = 60$) for packing fraction $\phi = 0.48$. The passive rings distribution is almost uniform throughout the box indicates that there is no clustering for $\text{Pe = 0}$.\\

\item \textbf{Movie\_S2} \\
\noindent Dynamic clustering of active ($\text{Pe = 50}$) rings ($n = 50, \, N = 60$) for packing fraction $\phi = 0.48$. Active rings start to form clusters and the cluster undergo degradation with time compared to the behavior observed in Movie\_S1. \\ 

\item \textbf{Movie\_S3} \\
\noindent Dynamic clustering of active ($\text{Pe = 50}$) rings ($n = 50, \, N = 60$) for packing fraction $\phi = 0.64$. Active rings undergo clustering followed by degradation with time compared to the behavior observed in Movie\_S2. Increasing $\phi$ facilitates clustering but the cluster size decreases while cluster number increases. \\

\item \textbf{Movie\_S4} \\
\noindent Dynamic clustering of active ($\text{Pe = 50}$) rings ($n = 50, \, N = 60$) for packing fraction $\phi = 0.72$. Active rings form smaller dense clusters and the cluster undergo degradation with time compared to the behavior observed in Movie\_S2. The cluster size is smaller compared to lower packing fraction (Movie\_S2 and Movie\_S3). At higher $\phi$, formation of large number of dense smaller clusters are favored.\\
  
\end{enumerate}

\clearpage
%
\end{document}